\documentstyle[12pt]{article}
\begin{document}

\def\ds{\displaystyle}
\def\beq{\begin{equation}}
\def\eeq{\end{equation}}
\def\bea{\begin{eqnarray}}
\def\eea{\end{eqnarray}}
\def\ve{\vert}
\def\vel{\left|}
\def\ver{\right|}
\def\nnb{\nonumber}
\def\ga{\left(}
\def\dr{\right)}
\def\aga{\left\{}
\def\adr{\right\}}
\def\lla{\left<}
\def\rra{\right>}
\def\rar{\rightarrow}
\def\nnb{\nonumber}
\def\la{\langle}
\def\ra{\rangle}
\def\ba{\begin{array}}
\def\ea{\end{array}}
\def\tr{\mbox{Tr}}
\def\ssp{{\Sigma^{*+}}}
\def\sso{{\Sigma^{*0}}}
\def\ssm{{\Sigma^{*-}}}
\def\xis0{{\Xi^{*0}}}
\def\xism{{\Xi^{*-}}}
\def\qs{\la \bar s s \ra}
\def\qu{\la \bar u u \ra}
\def\qd{\la \bar d d \ra}
\def\qq{\la \bar q q \ra}
\def\gGgG{\la g^2 G^2 \ra}
\def\q{\gamma_5 \not\!q}
\def\x{\gamma_5 \not\!x}
\def\g5{\gamma_5}
\def\sb{S_Q^{cf}}
\def\sd{S_d^{be}}
\def\su{S_u^{ad}}
\def\ss{S_s^{??}}
\def\sbp{{S}_Q^{'cf}}
\def\sdp{{S}_d^{'be}}
\def\sup{{S}_u^{'ad}}
\def\ssp{{S}_s^{'??}}
\def\sig{\sigma_{\mu \nu} \gamma_5 p^\mu q^\nu}
\def\fo{f_0(\frac{s_0}{M^2})}
\def\ffi{f_1(\frac{s_0}{M^2})}
\def\fii{f_2(\frac{s_0}{M^2})}
\def\O{{\cal O}}
\def\sl{{\Sigma^0 \Lambda}}
\def\es{\!\!\! &=& \!\!\!}
\def\ar{&+& \!\!\!}
\def\ek{&-& \!\!\!}
\def\cp{&\times& \!\!\!}
\def\se{\!\!\! &\simeq& \!\!\!}
\title{
         {\Large
                 {\bf
$\Sigma_Q \Lambda_Q$ transition 
magnetic moments in light cone QCD sum rules
                 }
         }
      }

\author{\vspace{1cm}\\
{\small T. M. Aliev$^a$ \thanks
{e-mail: taliev@metu.edu.tr}\,\,,
A. \"{O}zpineci$^b$ \thanks
{e-mail: ozpineci@ictp.trieste.it}\,\,,
M. Savc{\i}$^a$ \thanks
{e-mail: savci@metu.edu.tr}} \\
{\small a Physics Department, Middle East Technical University, 
06531 Ankara, Turkey}\\
{\small b  The Abdus Salam International Centre for Theoretical Physics,
I-34100, Trieste, Italy} }
\date{}

\begin{titlepage}
\maketitle
\thispagestyle{empty}

\begin{abstract}

Using the general form of $\Sigma_Q$ and $\Lambda_Q$ ($Q=b$ or $c$)
currents, $\Sigma_Q \Lambda_Q$ transition magnetic moments are
calculated in framework of the light cone QCD sum rules method. In this 
approach nonperturbative effects are described by the photon wave 
functions and only two--particle photon wave functions are taken into 
account. Our predictions on transition magnetic moments are 
$\mu_{\Sigma_c \Lambda_c} = - (1.5 \pm 0.4) \mu_N$ and
$\mu_{\Sigma_b \Lambda_b} = - (1.6 \pm 0.4) \mu_N$. A comparison of our
results with the ones existing in the literature is given. 

\end{abstract}

~~~PACS number(s): 11.55.Hx, 13.40.Em, 14.20.Lq, 14.20.Mr
\end{titlepage}

\section{Introduction}

Among all nonperturbative approaches, the QCD sum rules method \cite{R1},  
is an especially powerful one for calculating the fundamental parameters 
of hadrons. In this approach a deep connection is established between the 
hadronic parameters and the QCD vacuum via a few condensates. This method has been
applied very successfully to the various problems of hadron physics and is
discussed in detail in many review articles \cite{R2}--\cite{R7}. One of the
important static characteristic parameters of hadrons is their magnetic
moment. Nucleon magnetic moment was investigated in the framework of the
traditional QCD sum rules method in \cite{R8,R9}. Within this method, the $\Sigma
\Lambda$ transition magnetic moment was calculated in \cite{R10}.
Later this method was used in determining magnetic moments of baryons
containing heavy mesons \cite{R11}. In the present work our goal is to
calculate $\Sigma_Q \Lambda_Q$ ($Q=b$ or $c$) transition magnetic moment in 
the framework of the light cone QCD sum rules method (LCQSR) (more about this 
approach can be found in \cite{R7} and \cite{R12} and the references 
therein), which is an alternative approach to the traditional QCD sum rules 
method. Note that magnetic moments of the nucleons and decuplet baryons were
studied in \cite{R13,R14} using the LCQSR approach. Moreover, magnetic
moments of heavy $\Lambda_Q$ baryons and $\Sigma \Lambda$ transition 
magnetic moment were investigated in \cite{R15} and \cite{R16},
respectively, using the LCQSR method.

The paper is organized as follows: In section 2, the LCQSR for
$\Sigma_Q \Lambda_Q$ transition magnetic moment is derived. We present
numerical calculations and conclusion in section 3.       

\section{LCQSR for $\Sigma_Q \Lambda_Q$ transition magnetic moment}

In order to calculate $\Sigma_Q \Lambda_Q$ transition magnetic moment we
start by considering the following correlator function:
\bea
\label{pi1}
\Pi = i \int d^4 x \, e^{i p x} \lla 0 \ve T\{\eta_{\Lambda_Q}(x)
\bar \eta_{\Sigma_Q}(0) \} \ve 0 \rra_{ {\cal F}_{\alpha\beta}}~,   
\eea
where ${\cal F}_{\alpha\beta}$ is the external electromagnetic field,
$\eta_{\Sigma_Q}$ and $\eta_{\Lambda_Q}$ are the interpolating currents
with $\Sigma_Q$ and $\Lambda_Q$ quantum
numbers, respectively. It is well known that there is a continuum of 
choices for the baryon interpolating currents. The general form of
$\Sigma_Q$ and $\Lambda_Q$ currents can be written as \cite{R17,R18}
\bea                                                                  
\label{eta}                                                                
\eta_{\Lambda_Q} \es 2 \ga \eta_{\Lambda_1} + b \eta_{\Lambda_2}
\dr~, \nnb \\
\eta_{\Sigma_Q} \es 2 \ga \eta_{\Sigma_1} + b^\prime \eta_{\Sigma_2}
\dr~,
\eea
where $b$ and $b^\prime$ are arbitrary parameters and
\bea
\label{eta1}
\eta_{\Sigma_1} \es \frac{1}{\sqrt{2}} \epsilon_{abc}
\Big[(u_a^T C Q_b) \gamma_5 d_c + (d_a^T C Q_b) \gamma_5 u_c\Big]~, \\
\label{eta2}
\eta_{\Sigma_2} \es \frac{1}{\sqrt{2}} \epsilon_{abc}
\Big[(u_a^T C \gamma_5 Q_b) d_c + (d_a^T C \gamma_5 Q_b) u_c\Big] ~,\\
\label{eta3}
\eta_{\Lambda_1} \es \frac{1}{\sqrt{6}} \epsilon_{abc}
\Big[ 2 (u_a^T C d_b) \gamma_5 Q_c + (u_a^T C Q_b) \gamma_5 d_c - (d_a^T
C Q_b) \gamma_5 u_c \Big]~, \\
\label{eta4}
\eta_{\Lambda_2} \es \frac{1}{\sqrt{6}} \epsilon_{abc}
\Big[ 2 (u_a^T C \gamma_5 d_b) Q_c + (u_a^T C \gamma_5 Q_b) d_c -
(d_a^T C \gamma_5 Q_b) u_c \Big] ~,
\eea
with $a$, $b$, and $c$ being the color indices. Note that the Ioffe current 
corresponds to the choice $b=b^\prime=-1$.

Phenomenological part of the sum rules is obtained by inserting a complete
set of states with $\Sigma_Q$ and $\Lambda_Q$ quantum numbers between the
currents in Eq. (\ref{pi1}), which can be expressed as
\bea
\label{pi2}
\Pi \es \frac{\lla 0 \ve \eta_{\Sigma_Q} \ve \Sigma_{Q}(p_1) \rra}
{p_1^2 - m_{\Sigma_Q}^2} \lla \Sigma_{Q}(p_1) \ve \Lambda_{Q} (p_2) 
\rra_{ {\cal F}_{\alpha\beta}}
\frac{\lla \Lambda_{Q}(p_2) \ve \bar \eta_{\Lambda_Q} \ve 0 \rra}{p_2^2 -
m_{\Lambda_Q^2}^2} \nnb \\
\ar \sum_{h_i,H_i} 
\frac{\lla 0 \ve \eta_{\Sigma_Q} \ve h_i(p_1) \rra}
{p_1^2 - m_{h_i}^2} \lla h_i(p_1) \ve H_i (p_2) \rra_{
{\cal F}_{\alpha\beta}}
\frac{\lla H_i (p_2) \ve \bar \eta_{\Lambda_Q} \ve 0 \rra}{p_2^2 -
m_{H_i}^2}~,
\eea
where $p_2 = p_1 + q$, and $q$ is the photon momentum. The second term in Eq.
(\ref{pi2}) takes higher resonances and continuum contributions into account.

The coupling of currents with the corresponding 
baryon states is parametrized by the overlap amplitudes $\lambda_i$ which 
are defined as
\bea
\label{amp1}
\lla 0 \ve \eta_{\Sigma_Q} \ve \Sigma_Q \rra \es \lambda_{\Sigma_Q}
u_{\Sigma_Q} (p)~,\nnb \\
\lla 0 \ve \eta_{\Lambda_Q} \ve \Lambda_Q \rra \es \lambda_{\Lambda_Q}
u_{\Lambda_Q} (p)~.
\eea

It follows from Eq. (\ref{pi2}) that in calculating the
phenomenological part of the correlator, we need the expression
of the matrix element
$\lla \Sigma_{Q} (p_1) \ve \Lambda_{Q} (p_2) 
\rra_{ {\cal F}_{\alpha\beta}}$, which can be parametrized as
\bea
\label{amp2}
\lla \Sigma_Q(p_1) \ve \Lambda_Q(p_2) \rra_{ {\cal F}_{\alpha\beta}} \es
\bar u (p_1)
\left[ f_1 \gamma_\mu + i \, \frac{\sigma_{\mu\alpha} q^\alpha}
{m_{\Sigma_Q}+m_{\Lambda_Q}}
f_2 \right] u (p_2) \varepsilon^\mu~, \nnb \\
\es \bar u (p_1) \left[ (f_1 + f_2) \gamma_\mu + \frac{(p_1 + p_2)_\mu}
{m_{\Sigma_Q}+m_{\Lambda_Q}} f_2 \right] u (p_2) \varepsilon^\mu~,
\eea
where $f_1(q^2)$ and $f_2(q^2)$ are the form factors which are 
functions of $q^2 = (p_2 - p_1)^2$, and $\varepsilon^\mu$ is the 
polarization vector of photon.  
In order to calculate $\Sigma_Q\Lambda_Q$ transition magnetic
moment only the values of the form factors at $q^2=0$ are needed,
since the photon is a real one.
Using Eqs. (\ref{pi2})--(\ref{amp2}), for the hadronic representation of the
correlator function, we get
\bea
\label{pi3}
\Pi  = - \lambda_{\Sigma_Q} \lambda_{\Lambda_Q}
\varepsilon^\mu
\frac{\not\!p_1 + m_{\Sigma_Q}}{p_1^2 - m_{\Sigma_Q}^2}
\left[ (f_1 + f_2) \gamma_\mu + \frac{(p_1 +
p_2)_\mu}{m_{\Sigma_Q}+m_{\Lambda_Q}}
f_2 \right] \frac{\not\!p_2 + m_{\Lambda_Q}}{p_2^2 -   
m_{\Lambda_Q}^2}~.
\eea

Among a number of different structures which appear in the phenomenological 
part, we choose the structure
$\not\!p_1\!\!\not\!\varepsilon\!\!\not\!p_2$ that contains transition 
magnetic form factor
$f_1+f_2$. This form factor gives transition magnetic moment in units of 
$e\hbar/(m_{\Sigma_Q}+m_{\Lambda_Q})$ at $q^2=0$.
Choosing the above mentioned structure from the phenomenological part, we get
\bea
\label{pi4}
\Pi = - \lambda_{\Sigma_Q} \lambda_{\Lambda_Q} \frac{1}{p_1^2 -
m_{\Sigma_Q}^2}
\mu_{\Sigma_Q \Lambda_Q} \frac{1}{p_2^2 - m_{\Lambda_Q}^2}~,
\eea
where $\mu_{\Sigma_Q \Lambda_Q} = (f_1 + f_2) \ve_{q^2=0}$ is the transition
magnetic moment.

In order to construct sum rules, theoretical part of the
correlator function needs to be calculated. 
Calculation of the correlator function in QCD leads to the following result:

\bea
\label{cor}
\lefteqn{
\Pi(p,q) = - i \frac{2}{\sqrt{3}} \epsilon_{abc} \epsilon_{def} 
\int d^4x e^{i p x} \la \gamma(q) \ve } \nnb \\
&\phantom{+}& \!\!\! 2 		\g5 \sd \sup \sb \g5 
+2 b		\g5 \sd \g5 \sup \sb
+2 b'		\sd \sup \g5 \sb \g5 \nnb \\
\ar 2 b b'	\sd \g5 \sup \g5 \sb
+		\g5 \sd \g5 \tr \su \sbp 
+b		\g5 \sd \tr \su \g5 \sbp \nnb \\
\ar b'		\sd \g5 \tr \su \sbp \g5 
+b b'		\sd \tr \su \g5 \sbp \g5
- 		\g5 \sd \sbp \su \g5 \nnb \\
\ek b		\g5 \sd \g5 \sbp \su
-b'		\sd \sbp \g5 \su \g5
-b b'		\sd \g5 \sbp \g5 \su \nnb \\
\ek 2		\g5 \su \sdp \sb \g5 
-2b		\g5 \su \g5 \sdp \sb
-2 b'		\su \sdp \g5 \sb \g5 \nnb \\
\ek 2 b b'	\su \g5 \sdp \g5 \sb
+		\g5 \su \sbp \sd \g5 
+b		\g5 \su \g5 \sbp \sd \nnb \\
\ar b'		\su \sbp \g5 \sd \g5 
+b b'		\su \g5 \sbp \g5 \sd
-		\g5 \su \g5 \tr \sdp \sb \nnb \\
\ek b		\g5 \su \tr \g5 \sdp \sb
-b'		\su \g5 \tr \sdp \g5 \sb 
-b b'		\su \tr \g5 \sdp \g5 \sb \ve 0 \ra~,
\eea
where $S_q^\prime = C S_q^T C$ with $C$ being the charge conjugation 
operator, $T$ denotes transpose of the operator and $S_q$ is the quark 
propagator with the subindices referring to the corresponding quarks.   
                                
Theoretical part of the correlator contains two pieces: perturbative and 
nonperturbative. Perturbative part corresponds to the case when photon is 
radiated from the
freely propagating quarks. Its expression can be obtained by making the
following replacement in each one of the quark propagators and keeping the
other two as they are in Eq. (\ref{cor})
\bea
\label{sss}
S^{ab}_{\alpha \beta} \rightarrow -\frac{1}{2} \left( \int dy \, {\cal F}^{\mu \nu}
y_\nu S^{free} (x-y) \gamma_\mu
S^{free}(y) \right)^{ab}_{\alpha \beta}~,
\eea
where the Fock--Schwinger gauge $x_\mu A^\mu(x) = 0$ has been used 
and $S^{free}$ is the free quark operator. In $x$--representation 
the propagator of the free massive quark is
\bea
\label{fmq}  
S_Q^{free} = \frac{m_Q^2}{4 \pi^2} \frac{K_1 \ga m_Q \sqrt{-x^2}
\dr}{\sqrt{-x^2}} - i \, \frac{m_Q^2 \not\!x}{4 \pi^2 x^2} K_2
\ga m_Q \sqrt{-x^2} \dr~,
\eea
where $m_Q$ is the heavy quark mass and $K_i$ are Bessel functions.
The expression for massless propagator $S_q^{free}=i \not\!x/(2 \pi^2 x^4)$ 
can be obtained from Eq. (\ref{fmq}) by 
making use of the following expansions for Bessel functions 
\bea
K_1(x) &\sim& \frac{1}{x} + {\cal O} (x)~,\nnb \\
K_2(x) &\sim& \frac{2}{x^2} - \frac{1}{2} + {\cal O} (x^2)~, \nnb
\eea
and then formally setting $m_Q \rar 0$.

The nonperturbative contributions can be obtained from Eq. (\ref{cor}) by
making the following replacement in each one of the massless quark 
propagators:
\bea   
\label{rep}
S^{ab}_{\alpha \beta} \rar - \frac{1}{4} \bar q^a A_j q^b \ga A_j
\dr_{\alpha \beta}~,\nnb
\eea
where $A_j = \Big\{ 1,~\gamma_5,~\gamma_\alpha,~i\gamma_5 \gamma_\alpha,
~\sigma_{\alpha \beta} /\sqrt{2}\Big\}$ and sum over $A_j$ is implied,
and the other two propagators are the full propagators involving both
perturbative and nonperturbative contributions. In order to calculate
perturbative and nonperturbative contributions, the explicit expressions of
the heavy and light quark propagators in the presence of external 
field are needed.  

The complete light cone expansion of the propagator in external
field is presented in \cite{R19}. It receives contributions
from the nonlocal operators 
$\bar q G q$, $\bar q G G q$, $\bar q q \bar q q$,
where $G$ is the gluon field strength tensor. 
Here we consider operators with only one gluon
field and neglect terms with two gluons $\bar q GG q$, and four
quarks $\bar q q \bar q q$. Neglecting these terms
can be justified on the basis of an expansion in
conformal spin \cite{R20}.  In this approximation massive and massless 
quark propagators are given by the following expressions,
\bea
\label{hvy}
i S_Q (x) \es
i S_Q^{free} (x) - i g_s \int \frac{d^4 k}{(2\pi)^4} \, e^{-ikx} \int_0^1
dv \Bigg[\frac{\not\!k + m_Q}{( m_Q^2-k^2)^2} \, G^{\mu\nu}(vx)
\sigma_{\mu\nu} \nnb \\
\ar \frac{1}{m_Q^2-k^2}\, v x_\mu G^{\mu\nu} \gamma_\nu \Bigg]~, \\ \nnb \\
\nnb \\
\label{lig}
i S_q (x)  \es
i S_q^{free}(x) - \frac{\lla \bar q q \rra}{12} \ga 1 +
\frac{x^2 m_0^2}{16} \dr \nnb \\
\ek i \, g_s \int dv \, \Bigg[ \frac{\not\!x}{16 \pi^2 x^2}\, G^{\mu\nu}(vx)
\sigma_{\mu\nu} - \frac{i}{4  \pi^2 x^2}\, v x_\mu G^{\mu\nu} \gamma_\nu
\label{lgh}
\Bigg]~,
\eea
where $S_Q^{free} (x)$ and $S_q^{free} (x)$ are the free
propagators of the heavy and light quarks, respectively, and 
$m_0$ is defined from the relation 
\bea
\lla \bar q i g_s G_{\mu\nu} \sigma^{\mu\nu} q \rra = m_0^2 \qq~. \nnb
\eea
In Eq. (\ref{lig}) we neglect the operators with dimensions
larger than five, since their contributions are negligible.

As can be seen from Eqs. (\ref{cor})--(\ref{lig}), in order to calculate 
the theoretical part of the correlator function (\ref{pi1}), the matrix elements 
$\lla \gamma \vel \bar q A_i q \ver 0 \rra$ of the
non--local operators between photon and vacuum states are needed.
Up to twist--4 these matrix elements are defined in terms of the photon wave 
functions in the following way \cite{R20}--\cite{R22}:
\bea
\label{pwf}
\la \gamma (q) \ve \bar q \gamma_\alpha \gamma_5 q \ve 0 \ra \es \frac{f}{4}
e_q \epsilon_{\alpha \beta \rho \sigma} \varepsilon^\beta
q^\rho x^\sigma \int_0^1 du \, e^{i u qx} \psi(u)~, \nnb \\
\la \gamma (q) \ve \bar q \sigma_{\alpha \beta} q \ve 0 \ra \es
i e_q \qq \int_0^1 du \, e^{i u q x} \Bigg\{ (\varepsilon_\alpha q_\beta -
\varepsilon_\beta q_\alpha) \Big[ \chi \phi(u) + x^2 \Big(g_1(u) - g_2(u)\Big)
\Big]  \nnb \\
\ar \Big[ qx (\varepsilon_\alpha x_\beta - \varepsilon_\beta x_\alpha) +
\varepsilon x (x_\alpha q_\beta - x_\beta q_\alpha) \Big] g_2
(u) \Bigg\}~.
\eea
In Eq. (\ref{pwf}), $\chi$ is the magnetic susceptibility of the quark condensate, 
$e_q$ is the quark charge, $\phi(u)$ and $\psi(u)$
are the leading twist--2 photon wave functions, while $g_1(u)$ and $g_2(u)$
are the twist--4 functions.

Theoretical part of the correlator (\ref{pi1}) can be 
obtained by substituting photon wave functions and expressions for massive
and massless quark propagators into Eq. (\ref{cor}). 
Sum rules for $\Sigma_Q\Lambda_Q$ transition magnetic moment 
are obtained by equating the phenomenological and theoretical parts of the 
correlator (\ref{pi1}). Performing double Borel transformations on the
variables $p_1^2=p^2$ and $p_2^2=(p+q)^2$ on both sides of the
correlator, which suppress the continuum and higher state contributions
(see \cite{R14,R23,R24} for details), $\Sigma_Q\Lambda_Q$ transition magnetic 
moment can be expressed as


\bea
\label{mgm}
\lefteqn{
\sqrt{3} \lambda_{\Sigma_Q} \lambda_{\Lambda_Q} \mu_{\Sigma_Q \Lambda_Q} 
\,e^{-\ga m_{\Lambda_Q}^2/M_1^2 +  m_{\Sigma_Q}^2/M_2^2 \dr}=
(e_d-e_u) \Bigg\{ }\nnb \\
&&\!\!\!\frac{3}{16 \pi^4}  (2 + b + b' + 2 b b') M^6 \Psi(2,-1,m_Q^2/M^2) \nnb \\ 
\ar \frac{m_Q}{8 \pi^2} \qq (-3 b + b' + 2 b b') \chi \varphi(u_0) 
M^4 \Psi(2,0,m_Q^2/M^2) \nnb \\ 
\ek \frac{1}{8 \pi^2}  (2 + b + b' + 2 b b') f \psi (u_0) M^4 
\Psi(1,-1,m_Q^2/M^2) \nnb \\ 
\ek \frac{m_0^2}{9 M^2}  \qq^2 (-1 + b)(1 + b') \big[g_1(u_0) 
- g_2(u_0) \big] \nnb \\
\cp \Big\{ 3 F_1(m_Q^2/M^2) - \Big[ F_4(m_Q^2/M^2)-F_5(m_Q^2/M^2)\Big]
\Big\} \nnb \\ 
\ek \frac{m_Q}{48 M^2} m_0^2  \qq f \psi (u_0)\nnb \\
\cp \Big\{ (-2 -b' -b + 4 b b') \Big[ F_4(m_Q^2/M^2)-F_5(m_Q^2/M^2)\Big] + 
3 (-1 + b b') F_5(m_Q^2/M^2) \Big\} \nnb \\ 
\ek \frac{m_Q}{8 \pi^2}  \qq (-2 -b' -b + 4 b b') M^2 \Psi(1,0,m_Q^2/M^2)  \\ 
\ek \frac{m_Q}{2 \pi^2}  \qq (b' -3 b + 2 b b') (g_1(u_0) - 
g_2(u_0)) M^2 \Psi(1,0,m_Q^2/M^2) \nnb \\ 
\ek \frac{\qq^2}{3} (-1+b)(1+b') \chi \varphi(u_0) M^2 F_5(m_Q^2/M^2)
\nnb \\ 
\ar \frac{4}{3}  \qq^2 (-1 +b)(1+b') (g_1(u_0) - g_2(u_0)) 
F_4(m_Q^2/M^2) \nnb \\ 
\ar \frac{m_0^2}{36} \qq^2 (-1+b)(1+b') \chi \varphi(u_0) 
\Big[ 3 F_4(m_Q^2/M^2) - F_5(m_Q^2/M^2) \Big] \nnb \\ 
\ek \frac{m_Q}{12}  (2 + b + b' -4 b b') \qq f \psi (u_0) 
F_5(m_Q^2/M^2) \nnb \\ 
\ar \frac{m_Q}{32 \pi^2} m_0^2 \qq \left[ (-2 - b' - b + 4 b b') 
F_5(m_Q^2/M^2) + 3 (-1 + b b') \Psi(1,1,m_Q^2/M^2) \right]\Bigg\}~,\nnb 
\eea
where the functions $\Psi(\alpha,\beta,z)$ and $F_i(z)$ are defined as
\bea
\label{psi}
\Psi (\alpha,\beta,z) \es \frac{1}{\Gamma(\alpha)}
\int_1^\infty dt \, e^{-t z} \, t^{\beta-\alpha-1}
(t-1)^{\alpha-1}~,~~~~~~(\alpha > 0)~, \nnb \\
F_1 (z) \es z \ga z - 2 \dr e^{-z}~, \nnb \\
F_2 (z) \es  \ga z^2 - 4 z + 2 \dr e^{-z}~, \nnb \\
F_3 (z) \es \ga z^2 - 6 z + 6 \dr e^{-z}~, \nnb \\
F_4 (z) \es z e^{-z}~, \nnb \\
F_5 (z) \es e^{-z}~,\nnb
\eea
and
\bea
u_0 = \frac{M_2^2}{M_1^2+M_2^2}~,~~~~~~M^2=\frac{M_1^2 M_2^2}{M_1^2+M_2^2}~,
\nnb
\eea
with $M_1^2$ and $M_2^2$ being the Borel parameters. Since masses of the 
initial and final baryons are very close to each other, we will set 
$m_{\Sigma_Q}=m_{\Lambda_Q}$ and $M_1^2=M_2^2=2 M^2$, from which it follows 
that $u_0=1/2$. It should be noted that Borel transformations of 
Bessel functions are given in \cite{R25}. 

It follows from Eq. (\ref{mgm}) that determination of 
$\mu_{\Sigma_Q\Lambda_Q}$ transition magnetic moment requires a knowledge 
of the residues $\lambda_{\Sigma_Q}$ and $\lambda_{\Lambda_Q}$. These
residues can be determined from heavy baryons mass sum rules
(for the $\Lambda_b$ case the residue is calculated in \cite{R15})


\bea
\label{ml1}
\lefteqn{
m_{\Lambda_Q} \lambda_{\Lambda_Q}^2 \, e^{-m_{\Lambda_Q}^2/M^2}
= \frac{m_Q}{32 \pi^4} 
(-13 + 2 b + 11 b^2) M^6 \Psi(3,0,m_Q^2/M^2)} \nnb \\ 
\ar \frac{\qq}{12 \pi^2} (1 + 4 b - 5 b^2) M^4 \Psi(1,-1,m_Q^2/M^2) \nnb \\
\ek \frac{m_Q}{36 M^2} m_0^2 \qq^2 \Big\{3 (5 + 2 b + 5 b^2) 
\Big[ F_4(m_Q^2/M^2)-F_5(m_Q^2/M^2)\Big] \\
\ar (-1+b)^2 F_5(m_Q^2/M^2) \Big\} \nnb \\ 
\ar \frac{\qq}{96 \pi^2} m_0^2 M^2 \left[(-1-4 b + 5 b^2) 
F_5(m_Q^2/M^2) + (-5 + 4 b + b^2) \Psi(1,0,m_Q^2/M^2) \right] \nnb \\ 
\ar \frac{m_Q}{6} \qq^2 (5 + 2 b + 5 b^2) F_5(m_Q^2/M^2)~, \nnb \\ \nnb \\
\label{ml2}
\lefteqn{
\lambda_{\Lambda_Q}^2 \, e^{-m_{\Lambda_Q}^2/M^2}
= \frac{3}{32 \pi^4} (5 + 2 b + 5 b^2) M^6
\Psi(3,-1,m_Q^2/M^2)} \nnb \\
\ek \frac{m_0^2}{72 M^2} \qq^2 \left[ (-26 + 4 b + 22 b^2) F_4(m_Q^2/M^2) + 
(-1+b)^2 F_5(m_Q^2/M^2) \right] \nnb \\ 
\ar \frac{m_Q}{12 \pi^2} \qq (1 + 4 b - 5 b^2) M^2 \Psi(2,0,m_Q^2/M^2) \\ 
\ar \frac{\qq^2}{18} (-13 + 2 b + 11 b^2) F_5(m_Q^2/M^2) \nnb \\ 
\ar \frac{m_Q}{96 \pi^2} m_0^2 \qq \left[ (-1  - 4 b + 5 b^2)
\Psi(1,0,m_Q^2/M^2) + 
6 (-1+b^2) \Psi(2,1,m_Q^2/M^2) \right]~, \nnb \\ \nnb \\ 
\label{ms1}
\lefteqn{
m_{\Sigma_Q} \lambda_{\Sigma_Q}^2 \, e^{-m_{\Sigma_Q}^2/M^2}
= \frac{3 m_Q}{32 \pi^4} 
(1- {b'})^2 M^6 \Psi(3,0,m_Q^2/M^2)}  \nnb \\ 
\ar \frac{3}{4 \pi^2} \qq (1-{b'}^2) M^4 \Psi(1,-1,m_Q^2/M^2) \nnb \\ 
\ek \frac{m_Q}{12 M^2} m_0^2 \qq^2 \Big\{ (5 + 2 b' + 5 {b'}^2) 
\Big[ F_4(m_Q^2/M^2)-F_5(m_Q^2/M^2)\Big] \\
\ar (3 + 2 b' + 3 {b'}^2) F_5(m_Q^2/M^2) \Big\} \nnb \\ 
\ar \frac{m_0^2}{32 \pi^2} \qq (-1+{b'}^2) M^2 
\left[ 7 F_5(m_Q^2/M^2) - \Psi(1,0,m_Q^2/M^2) \right] \nnb \\ 
\ar \frac{m_Q}{6} \qq^2 (5 + 2 b' + 5 {b'}^2) F_5(m_Q^2/M^2)~, \nnb \\ \nnb \\
\label{ms2}
\lefteqn{
\lambda_{\Sigma_Q}^2 e^{-m_{\Sigma_Q}^2/M^2}
= \frac{3}{32 \pi^4} (5 + 2 b' + 5 {b'}^2) M^6
\Psi(3,-1,m_Q^2/M^2)} \nnb \\ 
\ek \frac{m_0^2}{24 M^2} \qq^2 (1-b')^2 \left[ 2 F_4
(m_Q^2/M^2) -
F_5(m_Q^2/M^2) \right] \nnb \\
\ek \frac{3 m_Q}{4 \pi^2} (-1 + {b'}^2) \qq M^2 \Psi(2,0,m_Q^2/M^2) \\ 
\ar \frac{\qq^2}{6} (1-b')^2 F_5(m_Q^2/M^2) \nnb \\ 
\ar \frac{m_Q}{32 \pi^2} m_0^2 \qq (-1 + {b'}^2) \left[7 \Psi(1,0,m_Q^2/M^2) 
+ 6 \Psi(2,1,m_Q^2/M^2) \right]~. \nnb 
\eea
Eqs. (\ref{ml1}) and (\ref{ms1}) correspond to the structure proportional  
to the unit operator, while  Eqs. (\ref{ml2}) and (\ref{ms2}) correspond to
the structure $\not\!p$. Finally, we remark that 
subtraction of the continuum contribution in Eqs. (\ref{mgm})--(\ref{ms2}) can
be performed with the help of the following replacement
\bea
\label{sub}
M^{2n} \Psi(\alpha,\beta,m_Q^2/M^2) \rightarrow  
\frac{1}{\Gamma(\alpha)} \frac{1}{\Gamma(n)} \int_{m_Q^2}^{s_0} ds
e^{-s/M^2} \int_1^{s/m_Q^2} dt (s - t m_Q^2)^{n-1} t^{\beta - \alpha -1}
(t-1)^{\alpha - 1}~,
\eea
for $\alpha > 0$ and $n > 0$.

\section{Numerical analysis}  

In this section we present our numerical calculation for
$\mu_{\Sigma_Q\Lambda_Q}$ transition magnetic moment. It follows from Eq.
(\ref{mgm}) that the main input parameters of the LCQSR are the photon wave 
functions. The photon wave functions which we use in the present work are 
given as \cite{R20,R23}:
\bea
\phi(u) \es 6 u (1-u)~,~~~~~\psi(u) = 1~, \nnb \\
g_1(u) \es - \frac{1}{8}(1-u)(3-u)~,~~~~~g_2(u) = -\frac{1}{4} (1-u)^2~.\nnb
\eea
The values of the other input parameters entering to the sum rules are:
 $\chi(1GeV)=-4.4~GeV^{-2}$ \cite{R26} (in \cite{R27} this
quantity is estimated to be $\chi=-3.3~GeV^{-2}$), 
$\qq(1 ~ GeV)=-(0.243)^3~GeV^3$, $m_0^2=(0.8\pm0.2)~GeV^2$ \cite{R22},
$m_c=1.3~GeV$, $m_b=4.8~GeV$, and $f=0.028~GeV^2$. Few word about the value of $f$ are in order. 
The analysis of the magnetic susceptibility $\chi$ by the QCD sum rules (see \cite{R26} and \cite{R27}) 
shows that at an accuracy of the order of $30\%$, the dominant contribution to this quantity comes
from $\rho$ meson at $\mu = 1~GeV^2$. It has been suggested that on these argument that the vector dominance
model works sufficiently well for electromagnetic properties of hadrons in the constant external field limit $q \rar 0$.
To the quoted accuracy, in this limit for the normalization constant is obtained \cite{R28}
$$
f \simeq \frac{e_u}{g_\rho} f_\rho m_\rho \simeq 0.028 \pm 0.009
$$
at $g_\rho = 5.5$ and $f_\rho = 0.2$.

The parameters $b$ and $b^\prime$ are completely arbitrary and the physical quantities should
not depend on the precise value of these parameters. To simplify further analysis,  we will 
set $b=b^\prime$. Since transition magnetic moment is a physical quantity 
it must be independent of the auxiliary parameters $b$, Borel mass square 
$M^2$ and continuum threshold $s_0$. So, our problem reduces to 
determining the respective regions for which $\mu_{\Sigma_Q\Lambda_Q}$ 
transition magnetic moment is independent of the above--mentioned 
parameters.

For this aim we prefer to consider the following three steps. In the
first step, we try to find the working region of $M^2$ where
$\mu_{\Sigma_Q\Lambda_Q}$ is independent of the Borel parameter $M^2$, 
at fixed values of $b$ and $s_0$. Along these lines, we present in 
Figs. (1) and (2) the dependence of $\mu_{\Sigma_b\Lambda_b}$ and 
$\mu_{\Sigma_c\Lambda_c}$ on $M^2$, respectively. From both figures we 
observe that, except $b=-1$ case, transition magnetic moments seem to be 
almost independent for the different choices of $b$ and $s_0$. The 
working region for $\mu_{\Sigma_b\Lambda_b}$ transition magnetic moment 
is $15~GeV^2 < M^2 < 30~GeV^2$, while it is  $2~GeV^2 \le M^2 \le 6~GeV^2$ 
for the $\mu_{\Sigma_c\Lambda_c}$ case. 

The next step is to determine the working region for the parameter $b$. 
For this purpose we use the fact that both mass sum rules for $\Lambda_Q$
and $\Sigma_Q$ should be positive. Note also that the mass of the $\Lambda_Q$ 
baryon can be obtained by dividing Eq. (\ref{ml1}) with Eq. (\ref{ml2}), 
and that of $\Sigma_Q$ can be obtained by dividing Eq. (\ref{ms1}) with 
Eq. (\ref{ms2}). In order to see whether this requirement is fulfilled or 
not, in Figs. (3) and (4) we present the dependence of the sum rules for 
the masses of the above--mentioned baryons on $\cos\theta$,
where $\theta$ is determined from the relation $\tan\theta=b$. It follows
from these figures that the working region of $b$, which guarantees the
positiveness of the sum rules, are in the intervals $-0.7 \le \cos\theta \le
+0.7$ for $\Lambda_b$ baryon and $-0.75 \le \cos\theta \le +0.7$ for $\Lambda_c$
baryon. Similar analysis for the $\Sigma_Q$ baryons leads to the following
result for the working region of the parameter $b$: $-0.7 \le \cos\theta
\le +0.7$ and $-0.8 \le \cos\theta \le +0.7$ for the $\Sigma_b$ and
$\Sigma_c$ baryons, respectively. So, the common working region for the
parameter $b$ for both cases is $-0.7 \le \cos\theta \le +0.7$.  

Having decided about the restriction of the parameter $b$, our third and final 
step is to determine the main objective of the present work, i.e.,
$\mu_{\Sigma_Q\Lambda_Q}$ transition magnetic moment. As before, we are
supposed to find a region for the parameter $b$ where transition
magnetic moment $\mu_{\Sigma_Q\Lambda_Q}$ is independent of its variation.
In the first step of our analysis we have decided on the working region of 
the Borel parameter $M^2$ for which $\mu_{\Sigma_Q\Lambda_Q}$ does not depend 
on its variation, and additionally, we have verified its insensitiveness to 
several different choices of the continuum threshold $s_0$. Fig. (5) presents 
the dependence of transition magnetic moment $\mu_{\Sigma_b\Lambda_b}$ 
(in units of the nucleon magneton $\mu_N$) on $\cos\theta$, at $M^2=25~GeV^2$ 
and at three fixed values of $s_0$. Similarly, in Fig. (6) we present the 
dependence of $\mu_{\Sigma_c\Lambda_c}$ on $\cos\theta$, at $M^2=4~GeV^2$ and 
at three fixed values of $s_0$. From both figures we deduce that
$\mu_{\Sigma_Q\Lambda_Q}$ is quite stable in the region $-0.5 \le \cos\theta
\le +0.08$, and can be said to be practically independent of the parameter 
$b$ and the continuum threshold $s_0$. It follows from all these
arguments that
\bea
\mu_{\Sigma_c\Lambda_c} \es - (1.5 \pm 0.4) \mu_N~,\nnb\\
\mu_{\Sigma_b\Lambda_b} \es - (1.6 \pm 0.4) \mu_N~,\nnb
\eea
where the uncertainty in the results can be mainly attributed to the
variation of the continuum threshold $s_0$, Borel parameter $M^2$ and the
twist--3 photon wave functions which are neglected, since we estimate that
their contribution to the transition magnetic moment is less than 5\%,
as well as to the uncertainties in the values of the parameters $f$, $\chi$, and
$m_0^2$.

Finally, we compare our results on $\mu_{\Sigma_Q \Lambda_Q}$ transition
magnetic moment with the ones predicted by other methods
in literature. Transition magnetic moment 
$\mu_{\Sigma_c\Lambda_c}$ is calculated in the LCQSR to
leading order in heavy quark effective theory which predicts
$\mu_{\Sigma_c\Lambda_c} = (1.0 \pm 0.2)\mu_N$ \cite{R29}. 
This result is calculated using Ioffe currents for
$\Sigma_c$ and $\Lambda_c$. When we compare our results with the results
given in \cite{R29} we see that the discrepancy between them is about
50\%. In our opinion this discrepancy can be attributed to the fact that the
result presented in \cite{R29} is calculated for the choice $b=-1$ (Ioffe
current), which is unphysical in our case. 

In summary, transition magnetic moments $\mu_{\Sigma_Q \Lambda_Q}$ are 
calculated in the framework of the LCQSR, using the general 
form of the interpolating currents for $\Sigma_Q$ and $\Lambda_Q$ mesons.
In our analysis, only two--particle photon wave functions are taken into account
while three--particle photon wave functions are neglected, since we estimate
that their contribution to the transition magnetic moments, as has already
been mentioned above, is less than 5\%.

\newpage

\section*{Figure captions}
{\bf Fig. (1)} The dependence of the transition magnetic moment 
$\mu_{\Sigma_b\Lambda_b}$ on
$M^2$ at two different values of the
continuum threshold $s_0=35~GeV^2$ and $s_0=45~GeV^2$, for several fixed 
values of the parameter $b$. Here in this figure and Figs. (2), (5) and (6)
the transition magnetic moments are given in
units of the nucleon magneton $\mu_N$. \\ \\
{\bf Fig. (2)} The dependence of the transition magnetic moment
$\mu_{\Sigma_c\Lambda_c}$ on
$M^2$ at two different values of the
continuum threshold $s_0=8~GeV^2$ and $s_0=12~GeV^2$, for several fixed
values of the parameter $b$.\\ \\
{\bf Fig. (3)} The dependence of the mass sum rule  
$m_{\Lambda_b}$ on $\cos \theta$, at $M^2=25~GeV^2$ and
at three different values of the
continuum threshold $s_0=35~GeV^2,~40~GeV^2$ and $45~GeV^2$.\\ \\
{\bf Fig. (4)} The dependence of the mass sum rule
$m_{\Lambda_c}$ on
$\cos \theta$, at $M^2=4~GeV^2$ and 
at three different values of the
continuum threshold $s_0=8~GeV^2,~10~GeV^2$ and $12~GeV^2$.\\ \\
{\bf Fig. (5)} The dependence of the transition magnetic moment
$\mu_{\Sigma_b\Lambda_b}$ on
$\cos \theta$, at the fixed value $M^2=25~GeV^2$ of the Borel parameter, 
and at three different values of the   
continuum threshold $s_0=35~GeV^2,~40~GeV^2$ and $45~GeV^2$.\\ \\
{\bf Fig. (6)} The dependence of the transition magnetic moment
$\mu_{\Sigma_c\Lambda_c}$ on
$\cos \theta$, at the fixed value $M^2=4~GeV^2$ of the Borel parameter, 
and at three different values of the
continuum threshold $s_0=8~GeV^2,~10~GeV^2$ and $12~GeV^2$.\\ \\
\newpage

\begin{figure}
\vskip 1.5 cm
    \includegraphics{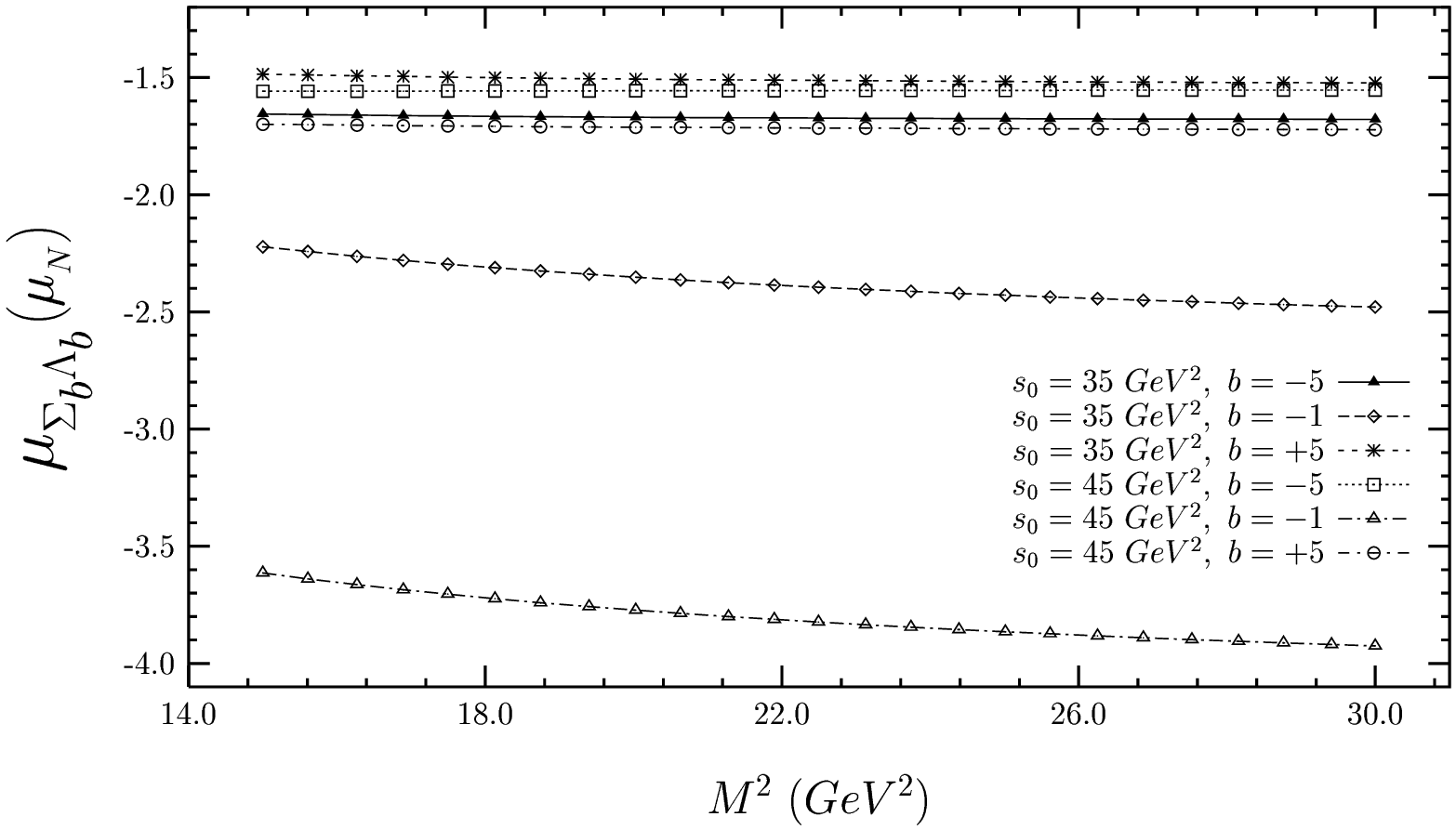}
\vskip 8.1cm
\caption{}
\end{figure}

\begin{figure}  
\vskip 1.5 cm
    \includegraphics{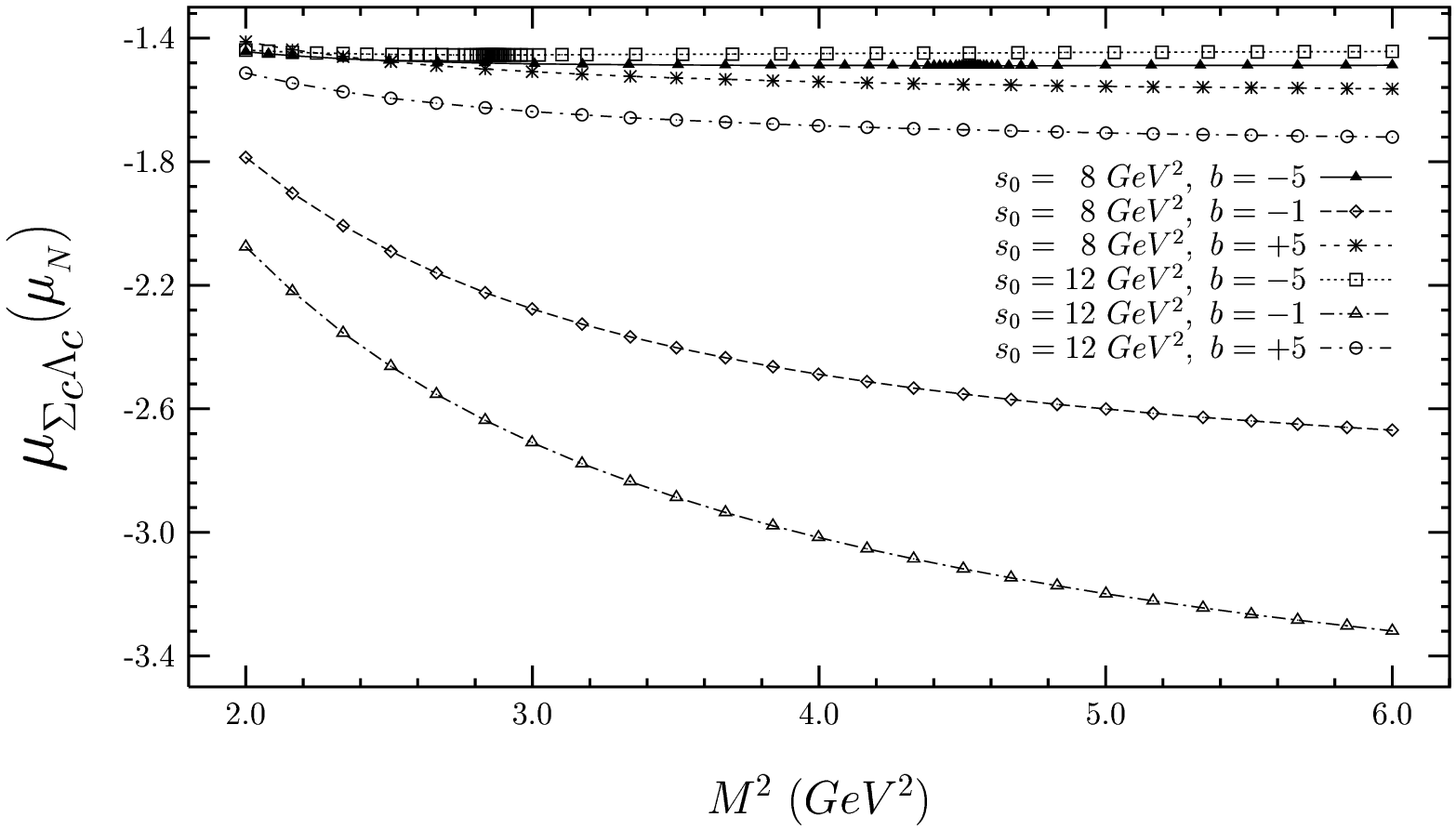}
\vskip 9. cm   
\caption{}
\end{figure}

\begin{figure}
\vskip 1.5 cm
    \includegraphics{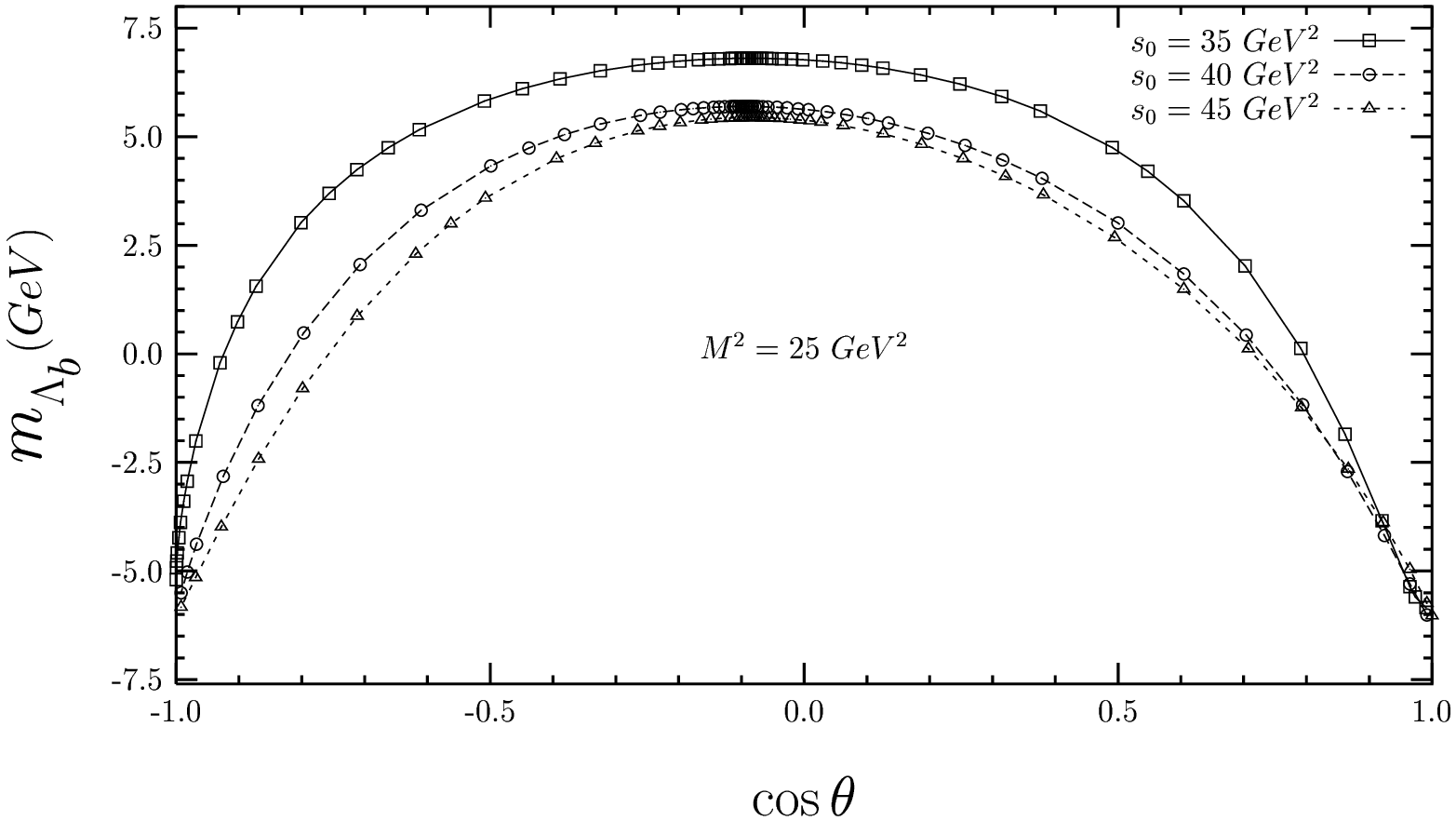}
\vskip 9. cm
\caption{}
\end{figure}

\begin{figure}
\vskip 1.5 cm
    \includegraphics{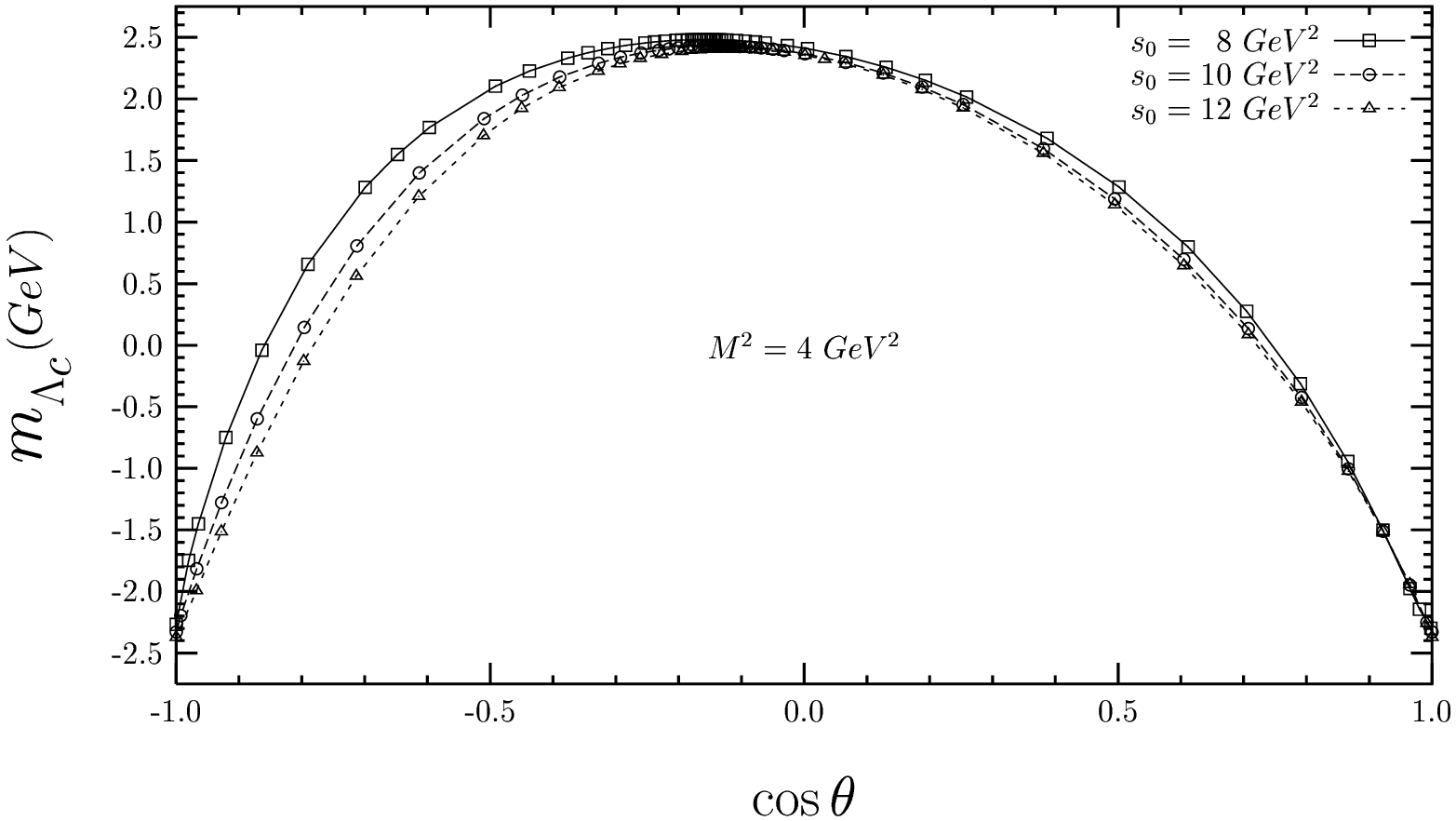}
\vskip 9. cm
\caption{}
\end{figure}

\begin{figure}
\vskip 1.5 cm
    \includegraphics{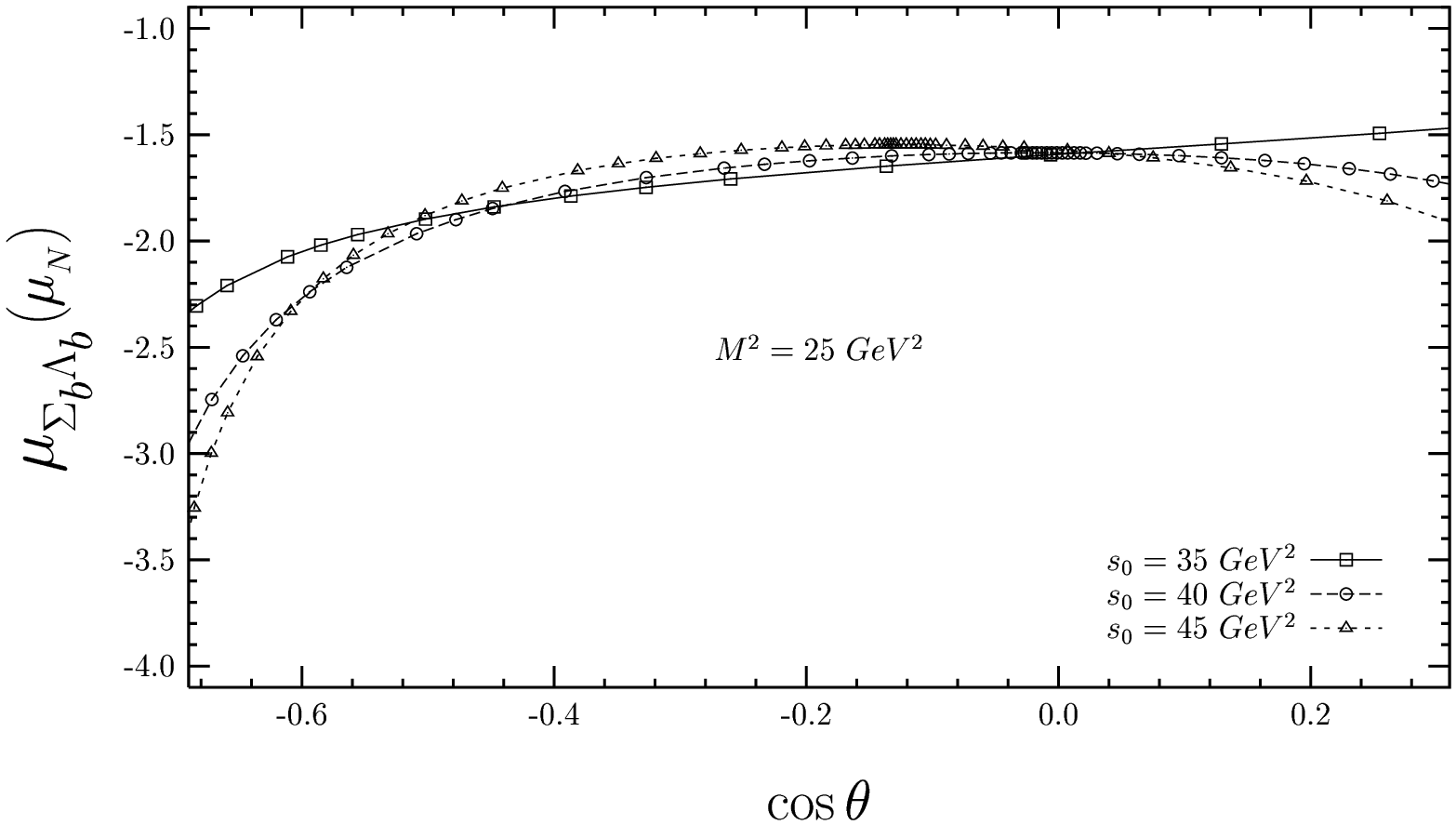}
\vskip 9. cm
\caption{}
\end{figure}

\begin{figure}
\vskip 1.5 cm
    \includegraphics{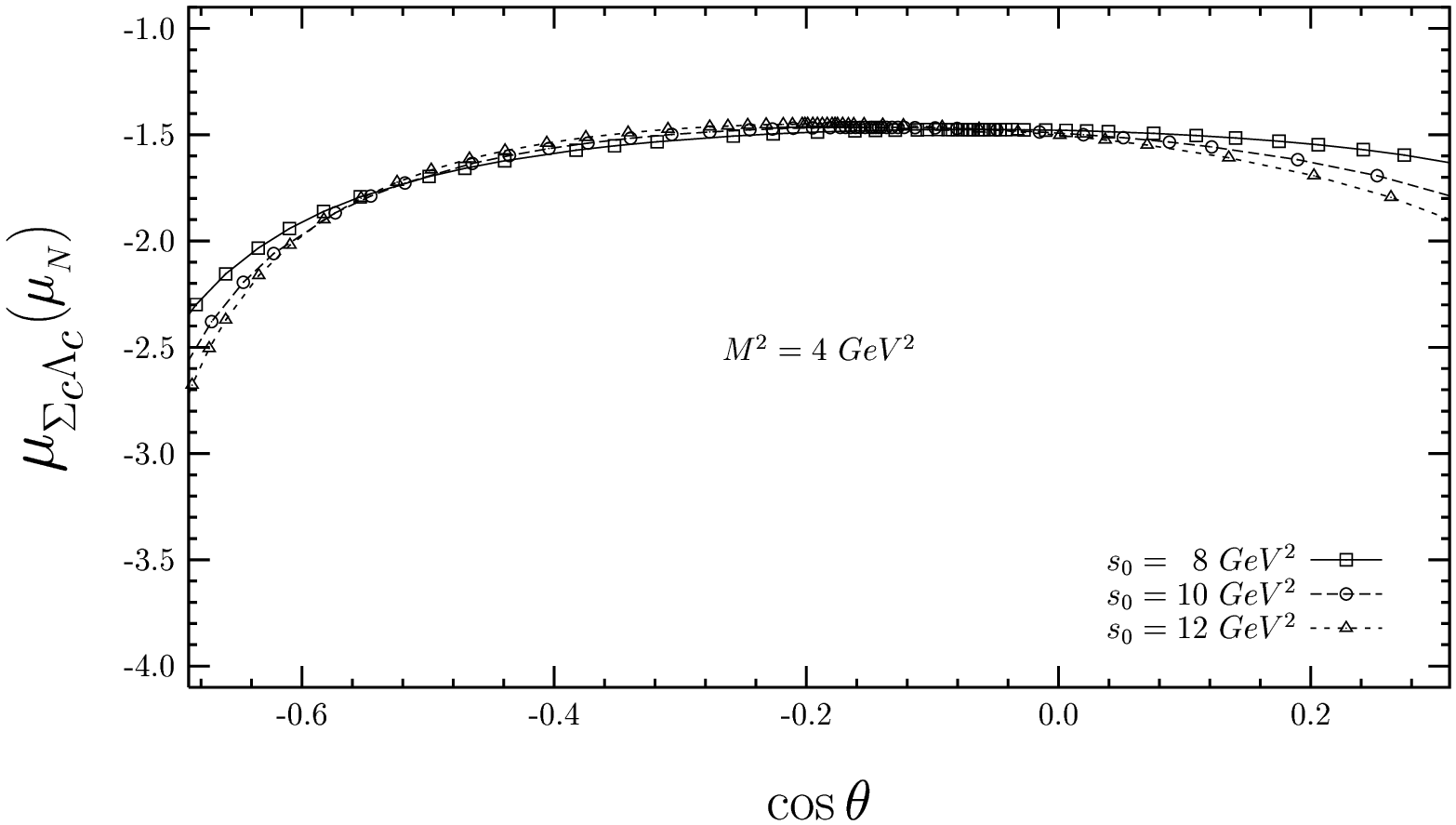}
\vskip 9. cm
\caption{}
\end{figure}

\end{document}